\documentclass[showpacs,preprintnumbers,prd,nofootinbib,floats,amssymb,floatfix]{revtex4}
\usepackage{amsmath}
\usepackage{graphicx}
\usepackage{amsxtra}
\usepackage{hyperref}
\usepackage{amsmath}
\usepackage{amstext}
\usepackage{blkarray} 
\begin{document}
\title{The Hamilton-Jacobi analysis for higher order Maxwell-Chern-Simons gauge theory}

\author{Alberto Escalante}  \email{aescalan@ifuap.buap.mx}
\author{ V\'ictor Alberto  Zavala-P\'erez}  \email{vzavala@ifuap.buap.mx}
 \affiliation{  Instituto de F{\'i}sica, Benem\'erita Universidad Aut\'onoma de Puebla. \\
 Apartado Postal J-48 72570, Puebla Pue., M\'exico, }
\begin{abstract}
Abstract. By using the Hamilton-Jacobi [$HJ$] framework the higher-order Maxwell-Chern-Simons theory is analyzed. The complete set of $HJ$ Hamiltonians and a generalized $HJ$ differential are reported, from which all symmetries of the theory are identified. In addition, we complete our study by performing the higher order Gitman-Lyakhovich-Tyutin [$GLT$] framework and compare the results of both formalisms.  
\end{abstract}
 \date{\today}
\preprint{}
\maketitle
\section{Introduction}
It is well known that higher-order theories are of interest in theoretical physics. In 1850 Ostrogradski developed works concerning the hamiltonian formalism for systems with higher derivatives \cite{1}. Since then, the research of systems such as generalizations of electrodynamics \cite{2, 3, 4, 5}, string  theory \cite{6}, and dark energy physics \cite{7, 8} has led to interesting results about gravity; where higher-order Lagrangians, which include quadratic products of the curvature tensor, have ensured the renormalization of these theories \cite{9, 10}. Moreover, when it comes to  a theory with gauge symmetries, the standard study of higher-order theories is done by using the so-called Ostrogradski-Dirac framework. \\
The Ostrogradski-Dirac framework is based on the extension of the phase space and on the choice of the fields and their temporal derivatives as canonical variables; the identification of the constraints is then performed as usual \cite{11}. Nonetheless, the classification of the constraints into first or second class is a difficult task. Alternative approaches can be employed, like the one developed by G\"uler \cite{F17} based on the identification of the constraints, called Hamiltonians, and the construction of a fundamental differential. These Hamiltonians  can be either involutive or non-involutive and are used to obtain the characteristic equations, the gauge symmetries, and the generalized $HJ$ brackets of the theory. Using this approach, the construction of the fundamental differential is straightforward and the identification of symmetries is, in general, more economical than other approaches \cite{F17,  F18, F19, F20, F21a, 12, 13}. \\
There is also a generalization of  Ostrogradski's framework, the so-called $GLT$ formalism \cite{14, 15}. Based on the introduction of the canonical momenta as  Lagrange multipliers, this framework allows one to reformulate the problem to one with only first-order time derivatives. Then, through a proper gauge fixing and making use of the Dirac brackets, the unphysical degrees of freedom can be removed.  At the end of the calculations the identification of the constraints is less complicated than in Ostrogadski's method. \\
In this paper we will analyze the higher order Maxwell-Chern-Simons gauge theory \cite{16, 17}.  We start with the G\"uler-$HJ$ approach,  we will introduce additional fields to reduce the problem to a first-order time derivative one. Due to this, non-physical degrees of freedom will appear, which are associated with non-involutive Hamiltonians; however, these will be removed with the introduction of the generalized $HJ$ brackets. In this manner, the identification of the Hamiltonians will be straightforward, we also extend the results reported in \cite{ 16, 17}. Incidentally, we report an alternative study beyond the Ostrogradski-Dirac framework in search of the best alternative for analyzing higher-order singular systems. Then, we will  finish our work  by performing a $GLT$ analysis. In fact, we will analyze the higher order Maxwell-Chern-Simons from two different perspectives and we will compare our results.   \\
The paper is organized as follows. In the Section II we develop the $HJ$ analysis for the higher order Maxwell-Chern-Simons gauge theory. We construct a fundamental differential, where the characteristics equations and all symmetries of the theory are found. Then,  we  reproduce and extend the results reported in \cite{16, 17}. In Section III the $GLT$ formalism is implemented; we  reduce the higher-order theory to a first-order one, then we identify all constraints of the theory and present a complete description of the Dirac algebra. 
\section{The Hamilton-Jacobi analysis}
The action under consideriation is given by \cite{16}
\begin{eqnarray}
\mathcal{L}=-\frac{1}{4} F^{\mu \nu} F_{\mu \nu}+\frac{\theta}{4} \epsilon^{\mu \nu \lambda} A_{\mu} F_{\nu \lambda}+\frac{1}{4 m} \epsilon^{\mu \nu \lambda}\left(\Box A_{\mu}\right) F_{\nu\lambda}, 
\label{Lag1}
\end{eqnarray}
where $A_\mu$ is the gauge potential, $F_{\mu \nu}$ is the curvature tensor, and $\epsilon^{\mu \nu \lambda}$ is the Levi-Cevita antisymmetric tensor. Throughout this paper we will use the following metric convection $\eta_{\mu \nu}= (-1, 1, 1)$, spacetime indices will be represented by  greek alphabet $\alpha, \beta = 0, 1, 2$, and space indices  by the Latin one $i, j, k =1, 2$. \\
By performing the $2+1$ decomposition we can write the action as
\begin{eqnarray} 
\mathcal{L}&=&\frac{1}{2}\dot{A}^{i}\dot{A}_{i} - \dot{A}_{i}\partial^{i}A_{0} - \frac{1}{2}\partial^{i}A^{0}\partial_{i}A_{0} - \frac{1}{4}F^{i j}F_{i j} + \frac{\theta}{2}\epsilon^{i j}A_{0}\partial_{i}A_{j} - \frac{\theta}{2}\epsilon^{i j}A_{i}\dot{A}_{j} + \frac{\theta}{2}\epsilon^{i j}A_{i}\partial_{j}A_{0} \nonumber \\[5pt]
&-& \frac{1}{2m}\epsilon^{i j}\ddot{A}_{0}\partial_{i}A_{j} + \frac{1}{2m}\epsilon^{i j}\nabla^{2}A_{0}\partial_{i} A_{j} + \frac{1}{2m}\epsilon^{i j}\ddot{A}_{i}\dot{A}_{j}  - \frac{1}{2m}\epsilon^{i j}\nabla^{2}A_{i}\dot{A}_{j} - \frac{1}{2m}\epsilon^{i j}\ddot{A}_{i}\partial_{j}A_{0} \nonumber \\[5pt]
&+& \frac{1}{2m}\epsilon^{i j}\nabla^{2}A_{i}\partial_{j} A_{0}, 
\label{Lag2}
\end{eqnarray}
For the purpose of analysis, we will write the Lagrangian (\ref{Lag2}) in a new fashion by  introducing the following variables $A_{\mu} \rightarrow \xi_{\mu}$,  $\dot{A}_{\mu} \rightarrow  v_{\mu}$. By doing this, the following constraints, given by  $\dot{\xi}_{\mu}- v_{\mu}=0$, will be added to the Lagrangian by means of new unphysical variables $\psi^{\mu}$. Thus, the Lagrangian takes the form
\begin{eqnarray}
\mathcal{L}&=& \frac{1}{2}v^{i}v_{i} - v_{i}\partial^{i}\xi_{0} - \frac{1}{2}\partial^{i}\xi^{0}\partial_{i}\xi_{0} - \frac{1}{4}F^{i j}F_{i j} + \frac{\theta}{2}\epsilon^{i j}\xi_{0}\partial_{i}\xi_{j} - \frac{\theta}{2}\epsilon^{i j}\xi_{i}v_{j} + \frac{\theta}{2}\epsilon^{i j}\xi_{i}\partial_{j}\xi_{0} \nonumber \\
&+& \frac{1}{2m}\epsilon^{i j}\left(-\dot{v}_{0} + \nabla^{2}\xi_{0}\right)\partial_{i}\xi_{j} - \frac{1}{2m}\epsilon^{i j}\left(-\dot{v}_{i} + \nabla^{2}\xi_{i}\right)v_{j} + \frac{1}{2m}\epsilon^{i j}\left(-\dot{v}_{i}+\nabla^{2}\xi_{i}\right)\partial_{j}\xi_{0} \nonumber \\[5pt]
&+& \psi^{0}\left(v_{0} - \dot{\xi}_{0}\right) + \psi^{i}\left(v_{i} - \dot{\xi}_{i}\right).     \label{Eq:eq1}
\end{eqnarray}
We can observe that the theory is now linear in the temporal derivatives and we can apply the $HJ$ analysis. From the definition of the momenta 
\begin{eqnarray*}
P^{\mu}=\frac{\partial \mathcal{L}}{\partial \dot{Q}_{\mu}},
\end{eqnarray*}
where $Q_{\mu} = (\xi_{0}, \xi_{i}, v_{0}, v_{i}, \psi_{0}, \psi_{i})$ are the canonical variables and $P^{\mu} = (\pi^{0}, \pi^{i}, \tilde{\pi}^{0}, \tilde{\pi}^{i}, p^{0}, p^{i})$ their corresponding momenta, we find the following Hamiltonians  \cite{F17,  F18, F19, F20, F21a, 12, 13}
\begin{eqnarray}
\Omega_{1}^{0} &=& \pi^{0} + \psi^{0} = 0, \nonumber \\
\Omega_{1}^{i} &=& \pi^{i} + \psi^{i} = 0, \nonumber \\
\Omega_{2}^{0} &=& \tilde{\pi}^{0} + \frac{1}{2m}\epsilon^{i j}\partial_{i}\xi_{j} = 0, \nonumber \\
\Omega_{2}^{i} &=& \tilde{\pi}^{i} - \frac{1}{2m}\epsilon^{i j}\left(v_{j} - \partial_{j}\xi_{0}\right) = 0, \nonumber \\
\Omega_{3}^{0} &=& p^{0} = 0, \nonumber \\[5pt]
\Omega_{3}^{i} &=& p^{i} = 0;
\label{const}
\end{eqnarray}
and the canonical Hamiltonian, given by 
\begin{eqnarray}
\mathcal{H} &=& \dot{\xi}_{\mu}\pi^{\mu} + \dot{v}_{\mu}\tilde{\pi}^{\mu} + \dot{\psi}_{\mu}p^{\mu} - \mathcal{L} \nonumber  \\[5pt]
 &=& -\frac{1}{2}v^{i}v_{i} + v_{i}\partial^{i}\xi_{0} + \frac{1}{2}\partial^{i}\xi^{0}\partial_{i}\xi_{0} + \frac{1}{4}F^{i j}F_{i j} - \frac{\theta}{2}\epsilon^{i j}\xi_{0}\partial_{i}\xi_{j} + \frac{\theta}{2}\epsilon^{i j}\xi_{i}v_{j} - \frac{\theta}{2}\epsilon^{i j}\xi_{i}\partial_{j}\xi_{0} \nonumber \\[5pt]
&+& \tilde{\pi}^{0}\nabla^{2}\xi_{0} + \tilde{\pi}^{i}\nabla^{2}\xi_{i} + \pi^{0}v_{0} + \pi^{i}v_{i}.
\end{eqnarray}
Thus, with the Hamiltonians identified, we construct the fundamental differential, which describes the evolution
of any function, say $F$, on the phase space \cite{F17, F18, F19, F20, F21a, 12, 13}
\begin{eqnarray}
dF &=& \int \Big[ \{F \;,\; \mathcal{H}\} dt^{0} + \{ F\;,\;\Omega_{1}^{0}\} d \omega^{1}_0 + \{F \;,\; \Omega_{1}^{i}\} d \omega^{1}_i + \{F \;,\; \Omega_{2}^{0} \} d \omega^{2}_0 + \{F \;,\; \Omega_{2}^{i} \} d \omega^{2}_ i \nonumber \\  
&+&  \{F \;,\; \Omega_{3}^{0} \} d \omega^{3}_ 0 + \{F \;,\; \Omega_{3}^{i} \} d \omega^{3}_i \Big] \;\; d^{2}y,
\end{eqnarray}
where $ \omega^{1}_0, \omega^{1}_i, \omega^{2}_0,  \omega^{2}_ i, \omega^{3}_ 0,  \omega^{3}_i $ are parameters associated with the Hamiltonians. To this end we separate the Hamiltonians into involutive and non-involutive. Involutive Hamiltonians are those whose Poisson brackets with all Hamiltonians, including themselves, vanish; otherwise, they are called non-involutive. These will be labeled by $\Gamma$ and $\Lambda$, respectively. The Poisson algebra between the Hamiltonians in (\ref{const}) is given by 
\begin{eqnarray}
\begin{matrix}
\begin{array}{ll}
\{\Omega_{1}^{0}(x) \;,\; \Omega_{2}^{i}(y)\} = -\frac{1}{2m}\epsilon^{i j}{\partial_{j}}\delta^{2}(x-y), & \{\Omega_{1}^{0}(x) \;,\; \Omega_{3}^{0}(y)\} = -\delta^{2}(x-y) \\[5pt]
\{\Omega_{1}^{i}(x) \;,\; \Omega_{2}^{0}(y)\} =  \frac{1}{2m}\epsilon^{i j}{\partial_{j}}\delta^{2}(x-y), & \{\Omega_{1}^{i}(x) \;,\; \Omega_{3}^{j}(y)\} = \eta^{i j}\delta^{2}(x-y)  \\[5pt]
\{\Omega_{2}^{i}(x) \;,\; \Omega_{2}^{j}(y)\} = -\frac{1}{m}\epsilon^{i j}\delta^{2}(x-y) 
\end{array}
\end{matrix}
,\end{eqnarray}
hence, we observe that all the Hamiltonians are non-involutive; particularly those related to the unphysical fields $\psi^{\mu}$ and their momenta $p_\mu$. The matrix composed of these Poisson brackets, namely 
\begin{eqnarray}
\Delta_{a b}=
\begin{pmatrix}
0 & 0 & 0 & -\frac{1}{2m}\epsilon^{j k}{\partial_{k}}& -1& 0 \\[5pt]
0 & 0 & \frac{1}{2m}\epsilon^{i j}{\partial_{j}} & 0 & 0 & \eta^{i j} \\[5pt]
0 & -\frac{1}{2m}\epsilon^{j k}{\partial_{k}}& 0 & 0 & 0 & 0 \\[5pt]
\frac{1}{2m}\epsilon^{i j}{\partial_{j}}& 0 & 0 & -\frac{1}{m}\epsilon^{i j} & 0 & 0 \\[5pt]
1 & 0 & 0 & 0 & 0 & 0 \\[5pt]
0 & -\eta^{i j} & 0 & 0 & 0 & 0 \\[5pt]
\end{pmatrix} \delta^2(x-y) \nonumber
,\end{eqnarray}
is not invertible, which means that the Hamiltonians are not independent. We will use the null vectors $\zeta$ of the matrix $\Delta_{a b}$ to identify the independent ones, such as it is done in a pure Dirac framework  \cite{18}
\begin{eqnarray}
\int d^{2} y \Delta^{\mu \nu} \zeta(y)_{\mu}=0.
\end{eqnarray}
Then a null vector is found, $\zeta_{\mu} = (0, 0, w, 0, 0, -\frac{1}{2m}\epsilon_{l j}\partial^jw)$, where $w$ is an arbitrary function. Contracting $\zeta_{\mu}$ with a vector composed of the non-involutive Hamiltonians $\Omega^{\mu} = (\Omega_{1}^{0}, \Omega_{1}^{i}, \Omega_{2}^{0}, \Omega_{2}^{i}, \Omega_{3}^{0}, \Omega_{3}^{i})$ yields a new Hamiltonian, given by 
\begin{eqnarray}
\zeta_{\mu}\Omega^{\mu}&=&0,  \nonumber \\
 \rightarrow \Gamma_1 &:& \tilde{\pi}^{0} + \frac{1}{2m}\epsilon^{i j}\partial_{i}\xi_{j} -\frac{1}{2m}\epsilon^{l j}\partial_{j}p_{l} = 0.
\label{invo1}
\end{eqnarray}
Since it's Poisson brackets with all other Hamiltonians (\ref{const}) vanishes, this new Hamiltonian is an involutive one. In this manner, the complete set of non-involutives Hamiltonians  is given by 
\begin{eqnarray}
\nonumber \Lambda_{1} &=& \pi^{0} + \psi^{0} = 0,     \label{Eq:eq24} \\ \nonumber 
\Lambda_2^ {i} &=& \pi^{i} + \psi^{i} = 0,     \label{Eq:eq25} \\ \nonumber 
\Lambda_3^{i} &=& \tilde{\pi}^{i} - \frac{1}{2m}\epsilon^{i j}\left(v_{j} - \partial_{j}\xi_{0}\right) = 0,     \label{Eq:eq26}  \\ \nonumber 
\Lambda_4 &=& p^{0} = 0,     \label{Eq:eq27} \\[5pt] 
\Lambda_5^{ i} &=& p^{i} = 0,.    \label{Eq:eq28} 
\end{eqnarray}
Thus, the new $\Delta_{a b}$ matrix, whose entries will be the Poisson brackets between the new non-involutive Hamiltonians (\ref{Eq:eq28}), takes the form 
\begin{eqnarray}
\Delta_{a b}=
\begin{pmatrix}
0 & 0 & -\frac{1}{2m}\epsilon^{j k}\partial_{k} & -1 & 0 \\[5pt]
0 & 0 & 0 & 0 & \eta^{i j} \\[5pt]
\frac{1}{2m}\epsilon^{i j}\partial_{j} & 0 & -\frac{1}{m}\epsilon^{i j} & 0 & 0 \\[5pt]
1 & 0 & 0 & 0 & 0 \\[5pt]
0 & -\eta^{i j} & 0 & 0 & 0 \\[5pt]
\end{pmatrix} \delta^{2}(x-y).
\end{eqnarray}
Which is found to not be singular; therefore it has an inverse, given by 
\begin{eqnarray}
\Delta_{ab}^{-1}(x, y) = 
\left(\begin{array}{ccccc}
0 & 0 & 0 & 1 & 0  \\[5pt]
0 & 0 & 0 & 0 & -\eta_{j l} \\[5pt]
0 & 0 & m\epsilon_{j l} & \frac{1}{2}{\partial_{j}}   & 0 \\[5pt]
-1 & 0 & -\frac{1}{2}{\partial_{l}} & 0 & 0 \\[5pt]
0 & \eta_{j l} & 0 & 0 & 0 \\[5pt]
\end{array}\right) \delta^{2}(x-y). \nonumber 
\end{eqnarray}
With this inverse matrix at hand we introduce the generalized brackets, defined as
\begin{eqnarray}
\{A(x), B(x^{\prime})\}^* = \{A(x), B(x^{\prime})\} - \int\int  \{A(x), \xi^{a}(y)\} \Delta^{{a b}^{-1}}(y, z) \{\xi^{b}(z), B(x^{\prime})\} \; d^{2}y \; d^{2}z
\label{gbra} 
,\end{eqnarray}
where $\xi^{\mu}$  represent the non-involutive Hamiltonians and $\Delta^{{a b}^{-1}}$ is the inverse of the matrix $\Delta_{a b}$, whose entries are the Poisson brackets between the non involutive Hamiltonians. Hence, by using the generalized brackets (\ref{gbra}) we can calculate the nontrivial ones between the phase space variables, these are 
\begin{eqnarray}
\begin{matrix}
\begin{array}{ll}
\{\xi_{\mu} \;,\; \pi^{\nu}\}^{*} = \delta_{\mu}^{\nu}\delta^{2}(x-y), & \{\xi_{\mu} \;,\; \psi_{\nu}\}^{*} = -\eta_{\mu \nu}\delta^{2}(x-y), \\[5pt]
\{\pi^{0} \;,\; v_{k}\}^{*} = \frac{1}{2}{\partial_{k}}\delta^{2}(x-y), & \{\pi^{0} \;,\; \tilde{\pi}^{k}\}^{*} = -\frac{1}{4m}\epsilon^{k l}{\partial_{l}}\delta^{2}(x-y), \\[5pt]
\{v_{0} \;,\; \tilde{\pi}^{0}\}^{*} = \delta^{2}(x-y), & \{v_{i} \;,\; v_{k}\}^{*}= m\epsilon_{i k} \delta^{2}(x-y), \\[5pt]
\{v_{i} \;,\; \tilde{\pi}^{k}\}^{*}= \frac{1}{2}\delta_{i}^{k}\delta^{2}(x-y), & \{v_{i} \;,\; \psi_{0}\}^{*} = \frac{1}{2}{\partial_{i}}\delta^{2}(x-y), \\[5pt]
\{\tilde{\pi}^{i} \;,\; \tilde{\pi}^{k}\}^{*} = \frac{1}{4m}\epsilon^{i k}\delta^{2}(x-y), & \{\tilde{\pi}^{i} \;,\; \psi_{0}\}^{*} =  -\frac{1}{4m}\epsilon^{i j}{\partial_{j}}\delta^{2}(x-y),  
\end{array}
\end{matrix}
\end{eqnarray} 
These generalized brackets will coincide with those of Dirac, which are calculated in the next section. In particular, we can observe that the generalized $HJ$ bracket between the Hamiltonian (\ref{invo1}) with itself gives 
\begin{eqnarray}
\{\Gamma_1(x) \;,\; \Gamma_1(y)\}^* =0
,\end{eqnarray}
confirming that $\Gamma_1$ is indeed involutive. In this manner, the introduction  of the $HJ$ brackets removes the non-involutive Hamiltonians and leaves us with a new fundamental differential, given by 
\begin{eqnarray}
dF&=& \int \Big[ \{F \;,\; \mathcal{H} (y)\}^* dt^{0} + \{ F\;,\;\Gamma_{1}(y)\}^* d \sigma^{1}  \Big]  \;\; d^{2}y.
\label{dfg}
\end{eqnarray}
By using the fundamental differential we have removed the unphysical degrees of freedom $\psi^{0}$ and  $\psi^{i}$, making the results in this section match those of the next section. In this regard, once the generalized brackets are introduced, we could perform the substitution of the fields $\psi 's$ by the momenta $\pi's$ in  the action (\ref{Eq:eq1}), the result would be that the $HJ$  and   $GLT$ actions are equivalent. In other approaches, see the ref. \cite{17}, the unphysical degrees of freedom are removed  until the end of the calculations, because the separation of the constraints into first and second class allows the introduction of the Dirac brackets; in contrast, in the $HJ$ framework, the elimination of unphysical degrees of freedom is more convenient. We also have to take into account the Frobenius integrability conditions, which ensure that system is integrable. Applying this conditions to the Hamiltonian $\Gamma_1$ the following Hamiltonian arises 
\begin{eqnarray}
d\Gamma_1(x)&=& \int \Big[ \{\Gamma_1(x) \;,\; \mathcal{H} (y)\}^* dt^{0} + \{ \Gamma_1(x) \;,\;\Gamma_{1}(y)\}^* d \sigma^{1}  \Big]  \;\; d^{2}y=0 \nonumber  \\ 
&\rightarrow& \Gamma_2  \equiv   \pi^{0} - \partial_{i}\tilde{\pi}^{i}=0, \nonumber
\label{invo2}
\end{eqnarray}
 we observe that, since $\{\Gamma_2(x), \Gamma_2(y) \}^*=\{\Gamma_2(x), \Gamma_1(y) \}^*=0$, $\Gamma_2$ is an involutive Hamiltonian. We add this new involutive Hamiltonian to the fundamental differential and then calculate it's integrability, obtaining a new involutive Hamiltonian
 \begin{eqnarray}
d\Gamma_2(x)&=& \int \Big[ \{\Gamma_2(x) \;,\; \mathcal{H} (y)\}^* dt^{0} + \{ \Gamma_2 (x)\;,\;\Gamma_{1}(y)\}^* d \sigma^{1}  +  \{ \Gamma_2(x)\;,\;\Gamma_{2}(y)\}^* d \sigma^{2}   \Big]  \;\; d^{2}y=0 \nonumber \\
&\rightarrow& \Gamma_3 \equiv  \partial_{i}\pi^{i}+ \frac{\theta}{2}\epsilon^{i j}\partial_{i}\xi_{j} +\frac{1}{2m}\epsilon^{i j}\nabla^{2}\partial_{i}\xi_{j} =0. \nonumber
\label{invo3}
\end{eqnarray}
No further Hamiltonians emerge from the integrability conditions of $\Gamma_{3}$. As a result, the complete set of involutive Hamiltonians is given by 
\begin{eqnarray}
\nonumber  \Gamma_{1} &=& \tilde{\pi}^{0} + \frac{1}{2m}\epsilon^{i j}\partial_{i}\xi_{j}=0,  \\ \nonumber 
\Gamma_2 &=& \pi^{0} - \partial_{i}\tilde{\pi}^{i}=0, \\ 
\Gamma_{3} &=&  \partial_{i}\pi^{i}+ \frac{\theta}{2}\epsilon^{i j}\partial_{i}\xi_{j} +\frac{1}{2m}\epsilon^{i j}\nabla^{2}\partial_{i}\xi_{j}=0.
\label{f1}
\end{eqnarray}
And the complete fundamental differential becomes
\begin{eqnarray}
dF&=& \!\! \int \Big[ \{F \;,\; \mathcal{H} (y)\}^* dt^{0} + \{ F\;,\;\Gamma_{1}(y)\}^* d \sigma^{1} + \{ F\;,\;\Gamma_{2}(y)\}^* d \sigma^{2} + \{ F\;,\;\Gamma_{3}(y)\}^* d \sigma^{3}  \Big]\;  d^{2}y
\label{dfg}
,\end{eqnarray}
where $\sigma^1, \sigma^2, \sigma^3$  are parameters associated to the Hamiltonians. 
Therefore, we have presented an alternative for studying higher-order theories in the context of $HJ$ theory, which is more  economical than those previously reported in the literature. With the fundamental  differential we can obtain the characteristic equations and then identify the symmetries, which is done in appendix A.   
\section{The Gitman-Lyakhovich-Tyutin framework }
In order to complete our analysis, the $GLT$ formalism will be performed. Starting with the Lagrangian (\ref{Lag2}) and following the formalism, we introduce the following variables \cite{14, 15}
\begin{eqnarray}
    v_{\mu} = \dot A_{\mu},
 \quad \quad 
    \beta_{\mu} = \dot v_{\mu},
\label{eq:var}
\end{eqnarray}
and their conjugated canonical momenta, satisfying 
\begin{eqnarray}
\nonumber 
\left\{A_{\mu}, \pi^{\nu}\right\} &=& \delta_{\mu}^{\nu}\delta^{2}(x-y),  \\ 
\left\{v_{\mu},  \tilde{\pi}^{\nu}\right\} &=& \delta_{\mu}^{\nu}\delta^{2}(x-y).
\label{bra}
\end{eqnarray}
Thus, the Lagrangian (\ref{Lag2}) can be written as 
\begin{eqnarray}
\mathcal{\tilde{L}} = \mathcal{L}+ \pi^{\mu}\left(\dot{A}_{\mu} - v_{\mu}\right) + \tilde{\pi}^{\mu}\left(\dot{v}_{\mu} - \beta_{\mu}\right), \nonumber
\end{eqnarray}
this is
\begin{eqnarray}
\nonumber
\mathcal{\tilde{L}}&=& \frac{1}{2}v^{i}v_{i} - v_{i}\partial^{i}A_{0} - \frac{1}{2}\partial^{i}A^{0}\partial_{i}A_{0} - \frac{1}{4}F^{i j}F_{i j} + \frac{\theta}{2}\epsilon^{i j}A_{0}\partial_{i}A_{j} - \frac{\theta}{2}\epsilon^{i j}A_{i}v_{j} + \frac{\theta}{2}\epsilon^{i j}A_{i}\partial_{j}A_{0}   \\  \nonumber 
&-& \frac{1}{2m}\epsilon^{i j}\beta_{0}\partial_{i}A_{j}  + \frac{1}{2m}\epsilon^{i j}\nabla^{2}A_{0}\partial_{i} A_{j} + \frac{1}{2m}\epsilon^{i j}\beta_{i}v_{j}  - \frac{1}{2m}\epsilon^{i j}\nabla^{2}A_{i}v_{j} 
- \frac{1}{2m}\epsilon^{i j}\beta_{i}\partial_{j}A_{0} \\ 
&+& \frac{1}{2m}\epsilon^{i j}\nabla^{2}A_{i}\partial_{j} A_{0}  + \pi^{\mu}\left(\dot{A}_{\mu} - v_{\mu}\right) + \tilde{\pi}^{\mu}\left(\dot{v}_{\mu} - \beta_{\mu}\right).
\label{Lag3}
\end{eqnarray}
We can observe that the theory is now first-order in the time derivatives, as well as that the canonical momenta have been introduced from the beginning. It is worth mentioning that the introduction of the momenta allows us to more easily identify the constraints compared to Ostrogradski's  formalism. In fact, in the $GLT$ framework it is not necessary to introduce a generalized canonical momenta for the higher-order time derivatives of the fields, as is done in Ostrogradski's framework \cite{14, 15, 16, 17}. Subsequently, the canonical Hamiltonian is given as usual 
\begin{eqnarray}
\mathcal{H} &=&  \dot{A}_{\mu }\pi^{\mu }  +  \dot{v}_{\mu }\tilde{\pi}^{\mu } - \mathcal{\tilde{L}}, \nonumber \\
 &=& v_{0}\pi^{0} + v_{i}\pi^{i}  + \beta_{0}\tilde{\pi}^{0} + \beta_{i}\tilde{\pi}^{i} - \frac{1}{2}v^{i}v_{i} + v_{i}\partial^{i}A_{0} + \frac{1}{2}\partial^{i}A^{0}\partial_{i}A_{0} + \frac{1}{4}F^{i j}F_{i j} - \frac{\theta}{2}\epsilon^{i j}A_{0}\partial_{i}A_{j} \nonumber  \\
&+& \frac{\theta}{2}\epsilon^{i j}A_{i}v_{j} - \frac{\theta}{2}\epsilon^{i j}A_{i}\partial_{j}A_{0} + \frac{1}{2m}\epsilon^{i j}\beta_{0}\partial_{i}A_{j} - \frac{1}{2m}\epsilon^{i j}\nabla^{2}A_{0}\partial_{i} A_{j} - \frac{1}{2m}\epsilon^{i j}\beta_{i}v_{j} + \frac{1}{2m}\epsilon^{i j}\nabla^{2}A_{i}v_{j} \nonumber \\
&+& \frac{1}{2m}\epsilon^{i j}\beta_{i}\partial_{j}A_{0} - \frac{1}{2m}\epsilon^{i j}\nabla^{2}A_{i}\partial_{j} A_{0}.
 \label{Ham1}
\end{eqnarray}
Thus, the primary constraints (called $\phi$) are given by \cite{14, 15}
\begin{eqnarray}
\phi^{3} &=& \frac{\partial\mathcal{L}}{\partial \beta_{0}} - \tilde{\pi}^{0} = -\frac{1}{2m}\epsilon^{i j}\partial_{i}A_{j} - \tilde{\pi}^{0}\approx 0,    \label{Eq:RestriccionGLT3} \\[5pt]
\phi^{i} &=& \frac{\partial\mathcal{L}}{\partial \beta_{i}} - \tilde{\pi}^{i} = \frac{1}{2m}\epsilon^{i j}v_{j} - \frac{1}{2m}\epsilon^{i j}\partial_{j}A_{0} - \tilde{\pi}^{i} \approx0.    \label{Eq:RestriccionGLTi}
\end{eqnarray}
They coincide only  with $\Omega_{2}^{0}$, and $\Omega_{2}^{i}$ from equation \ref{const}. Their algebra is given by 
\begin{eqnarray}
\nonumber \{\phi^{3} , \phi^{i}\} &=& 0, \\ 
\{\phi^{i} , \phi^{j}\} &=& - \frac{1}{m}\epsilon^{i j}\delta^{2}(x-y).
 \label{Eq:Poisson00}
\end{eqnarray}
At this point, it is important to comment the differences between the $HJ$ formalism and the GLT formulation. On one hand, in  $GLT$'s formulation we must  identify future constraints through consistency, then perform the classification of the constraints into first and second class, then Dirac's brackets are introduced and second class constraints can be taken strongly as zero. Only at the end of the calculations we can  compare both formalisms. On the other hand, in the $HJ$ scheme the generalized brackets, which has an equivalent construction just as  the Dirac ones, are introduced from the beginning. At the end of the calculations one ends up only with involutive Hamiltonians; which will agree with the set of first-class constraints of the $GLT$ formalism. \\
We continue with the classification of the constraints. By using the primary constraints we introduce the primary Hamiltonian 
\begin{eqnarray}
\mathcal{H}'= \mathcal{H}+\lambda_{3}\phi^{3}+\lambda_{i}\phi^{i}, 
\end{eqnarray}
where $\lambda_{3}$ and $\lambda_{i}$ are Lagrange multipliers, thus, by using (\ref{Eq:Poisson00}) and by requiring consistency of the primary constraints we obtain a secondary constraint
\begin{eqnarray}
\nonumber \chi^{0}: \dot{\phi}^{3} &=& \{\phi^{3} \;,\;\mathcal{H}'\} \\
&=& \pi^{0}-\frac{1}{2m}\epsilon^{i j}\partial_{i}v_{j} \approx 0
,\end{eqnarray}
Consistency of  $\phi^i$  provides a relation between the Lagrange multipliers, this is 
\begin{eqnarray}
\nonumber \dot{\phi}^{i} &=& \{\phi^{i} \;,\;\mathcal{H}'\} \\
&=&-\epsilon^{i j} \lambda_{j} + \epsilon^{i j}\beta_{j} -\frac{1}{2}\epsilon^{i j}\partial_{j}v_{0} + m\pi^{i} - mv^{i} + m\partial^{i}A_{0} - \frac{\theta m}{2}\epsilon^{i j}A_{j} - \frac{1}{2}\epsilon^{i j}\nabla^{2}A_{j} \approx0.
\label{12a}
\end{eqnarray}
Now, by demanding consistency of the secondary constraint $\chi^0$ we find 
\begin{eqnarray}
\nonumber \dot{\chi^0}&=& \{\chi^{0} \;,\;\mathcal{H}'\} \\
&=&-\epsilon^{i j}\partial_{i}\lambda_{j} + \epsilon^{i j}\partial_{i}\beta_{j} - m\partial^{i}v_{i} + m\nabla^{2}A^{0} - \theta m\epsilon^{i j}\partial_{i}A_{j} - \epsilon^{i j}\nabla^{2}\partial_{i}A_{j}\approx0,
     \label{Eq:consist-chi0-GLT}
\end{eqnarray}
which also contains relations between the Lagrange multipliers. Furthermore,  from (\ref{12a}) and (\ref{Eq:consist-chi0-GLT}) we can eliminate the Lagrange multipliers to obtain yet another secondary constraint
\begin{eqnarray}
\chi^{1} = \partial_{i}\pi^{i} + \frac{\theta }{2}\epsilon^{i j}\partial_{i}A_{j} + \frac{1}{2m}\epsilon^{i j}\nabla^{2}\partial_{i}A_{j}\approx0
     \label{Eq:Chi5-GLT}
,\end{eqnarray}
From consistency of $\chi^1$ no more constraints are found. In this manner, the complete set of $GLT$ constrains is given by
\begin{eqnarray}
\nonumber \phi^{3} &=& \tilde{\pi}^{0}+\frac{1}{2m}\epsilon^{i j}\partial_{i}A_{j} \approx 0, \nonumber \\[5pt]
\phi^{i} &=& \tilde{\pi}^{i} - \frac{1}{2m}\epsilon^{i j}v_{j} + \frac{1}{2m}\epsilon^{i j}\partial_{j}A_{0} \approx0, \nonumber    \\[5pt]
\chi^{0} &=&  \pi^{0} -\frac{1}{2m}\epsilon^{i j}\partial_{i}v_{j} \approx 0, \nonumber \\[5pt]
\chi^{1} &=& \partial_{i}\pi^{i} + \frac{\theta }{2}\epsilon^{i j}\partial_{i}A_{j} + \frac{1}{2m}\epsilon^{i j}\nabla^{2}\partial_{i}A_{j}\approx0. 
\label{fullcons}
\end{eqnarray}
Notice that $\phi^{i}$ is actually two constraints, so there are five in total. To separate them into first and second class we calculate the $5 \times 5$ matrix whose entries are the Poisson brackets between all constraints. This, in compact form, is
\begin{eqnarray}
\boldsymbol{A} = 
\begin{pmatrix}
 -\frac{1}{m}\epsilon^{i j} &       0       & \frac{1}{m}\epsilon^{i j}\partial_{j} &       0        \\[8pt]
       0       &       0       &       0       &       0        \\[8pt]
 -\frac{1}{m}\epsilon^{j i}\partial_{i}     &       0       &       0       &        0       \\[8pt]
       0       &       0       &       0       &       0        
\end{pmatrix} \delta^{2}(x-y)
     \label{Eq:matrixPoissonGLT}
,\end{eqnarray}
we observe that  this matrix   has a rank=2 and  3 null vectors, this means that there will be two second class constraints and three first class ones \cite{18}. The contraction of the null vectors with the constraints (\ref{fullcons}) allows us  identify the  following first class constraints
\begin{eqnarray}
\nonumber \gamma^{1} &=& \tilde{\pi}^{0}+ \frac{1}{2m}\epsilon^{i j}\partial_{i}A_{j},    \nonumber \\[5pt]
\gamma^{2} &=& \pi^{0}  -\partial_{i}\tilde{\pi}^{i},   \nonumber \\[5pt]
\gamma^{3} &=& \partial_{i}\pi^{i} + \frac{\theta }{2}\epsilon^{i j}\partial_{i}A_{j} + \frac{1}{2m}\epsilon^{i j}\nabla^{2}\partial_{i}A_{j}
\label{first}
,\end{eqnarray}
e.g. one null vector is given by $\tilde{v}= (0, \partial_i w, w, 0)$, and from the contraction with (\ref{fullcons}) we obtain $\gamma^2$.  We observe that the constraints (\ref{first}) coincide with the Hamiltonians (\ref{f1}) obtained with $HJ$ framework in the previous section. The two second class constraints are
\begin{eqnarray}
\xi^{i} = \tilde{\pi}^{i}- \frac{1}{2m}\epsilon^{i j}v_{j} + \frac{1}{2m}\epsilon^{i j}\partial_{j}A_{0}, 
\label{second}
\end{eqnarray}
these constraints are removed from the beginning in $HJ$ approach, in this sense the $HJ$ is more economical.  It is worth commenting that the constraints have been obtained in consistent form by using the ideas presented in \cite{18} and  it is not necessary to fix them by hand such as has been done previously in the literature \cite{17}. With the identification of the correct constraints, we can carry out the counting of physical degrees of freedom as follows: there are $12$ canonical variables $\left\{A_{\mu}, \pi^{\nu}\right\},  \left\{v_{\mu}, \tilde{\pi}^{\nu}\right\}$, three first class constraints $( \gamma^1, \gamma^2, \gamma^3)$ and two second class constraints $(\xi^i)$, therefore, there are two physical  degrees of freedom, as expected \cite{16}.\\
The second class constraints can be removed by means of the Dirac bracket
\begin{eqnarray}
\{A(x), B(x^{\prime})\}_{D} = \{A(x), B(x^{\prime})\} - \int\int  \{A(x), \xi^{a}(y)\} \Delta_{a b}(y, z) \{\xi^{b}(z), B(x^{\prime})\} \; d^{2}y \; d^{2}z
,\end{eqnarray}
where $\Delta_{a b}$ is the inverse of $\Delta^{a b}$, which consists of Poisson brackets among the second class constraints: $\Delta^{a b}=\{\xi^{a}, \xi^{b}\}$. This $2\times2$ matrix being as follows

\begin{eqnarray}
\Delta_{i j}(x, y) = m\epsilon_{i j}\delta^{2}(x-y)
\end{eqnarray}

 This results in the following non-trivial Dirac's brackets
\begin{eqnarray}
\begin{matrix}
\begin{array}{ll}
\{A_{0} \;,\; \pi^{0}\}_{D} = \delta^{2}(x-y), & \{A_{i} \;,\; \pi^{j}\}_{D} = \delta_{i}^{j}\delta^{2}(x-y), \\[5pt]
\{\pi^{0} \;,\; v_{i}\}_{D} = \frac{1}{2}\partial_{i}\delta^{2}(x-y), & \{\pi^{0} \;,\; \tilde{\pi}^{i}\}_{D}= -\frac{1}{4m}\epsilon^{i j}\partial_{j}\delta^{2}(x-y), \\[5pt]
\{v_{0} \;,\; \tilde{\pi}^{0}\}_{D} = \delta^{2}(x-y), & \{v_{i} \;,\; \tilde{\pi}^{j}\}_{D} = \frac{1}{2}\delta_{i}^{j} \delta^{2}(x-y),  \\[5pt]
\{v_{i} \;,\; v_{j}\}_{D} = m \epsilon_{i j} \delta^{2}(x-y), & \{\tilde{\pi}^{i} \;,\; \tilde{\pi}^{j}\}_{D} = \frac{1}{4m} \epsilon^{i j} \delta^{2}(x-y). \\[5pt]
\end{array}
\end{matrix}
\label{DiracGLT}
\end{eqnarray}
Using these we see that the constraints $\gamma^{1}$, $\gamma^{2}$, and $\gamma^{3}$ are still first class. 
We will now fix the gauge in order to remove  all  first class constraints, turning them into second class. It is important to comment that the gauge-fxing condition  removes the redundant degrees of freedom \cite{11}.  Demanding consistency of the Coulomb gauge $\gamma^{4} = \partial_{i}A^{i}$ results in
\begin{eqnarray}
\gamma^5: =\{\partial_{i}A^{i} \;,\; H\}_{D} = \partial_{i}v^{i}.
\end{eqnarray}
Demanding consistency of $\gamma^{5}$ yields $\gamma^{6}$
\begin{eqnarray}
\gamma^{6}: = \{\partial_{i}v^{i} \;,\; H\}_{D} = \frac{1}{2}\nabla^{2}v_{0} + m\epsilon^{i j}\partial_{i}\pi_{j} - m\epsilon^{i j}\partial_{i}v_{j},  
\end{eqnarray}
preservation in time of $\gamma^6$ gives no new constraints. Below we present the nontrivial  brackets among all  constraints
\begin{eqnarray}
\{\gamma^{4} \;,\; \gamma^{3}\}_{D} &=&  -\nabla^{2}\delta^{2}(x-y), \nonumber \\[5pt]
\{\gamma^{5} \;,\; \gamma^{2}\}_{D} &=& \nabla^{2}\delta^{2}(x-y), \nonumber \\[5pt]
\{\gamma^{6} \;,\; \gamma^{1}\}_{D} &=& \nabla^{2}\delta^{2}(x-y), \nonumber \\[5pt]
\{\gamma^{6} \;,\; \gamma^{3}\}_{D} &=& \frac{\theta m}{2}\nabla^{2}\delta^{2}(x-y) + \frac{1}{2}\nabla^{4}\delta^{2}(x-y), \nonumber \\[5pt]
\{\gamma^{5} \;,\; \gamma^{6}\}_{D} &=& m^{2}\nabla^{2}\delta^{2}(x - y). 
\label{alge}
\end{eqnarray}
Since, as can be easily seen, $\gamma_1,.., \gamma_6$ are all second class constraints, a new Dirac bracket can be introduced. In fact, by using (\ref{alge}) the new Dirac's brackets, say $\{, \}_{D_2}$, are given by
\begin{eqnarray}
\begin{matrix}
\begin{array}{ll}
\{A_{0} \;,\; v_{0}\}_{D_{2}} = m^{2}\frac{1}{\nabla^{2}}\delta^{2}(x-y), & \{A_{0} \;,\; v_{i}\}_{D_{2}} = m\epsilon_{i j}\frac{\partial^{j}}{\nabla^{2}}\delta^{2}(x-y), \\[5pt]
\{A_{0} \;,\; \pi^{0}\}_{D_{2}} = \delta^{2}(x-y), & \{A_{0} \;,\; \pi^{i}\}_{D_{2}} = -\frac{m}{2}\epsilon^{i j}\frac{\partial_{j}}{\nabla^{2}}\delta^{2}(x- y), \\[5pt]
\{A_{0} \;,\; \tilde{\pi}^{i}\}_{D_{2}} = -\frac{1}{2}\frac{\partial^{i}}{\nabla^{2}}\delta^{2}(x-y), & \{A_{i} \;,\; v_{0}\}_{D_{2}} = -m\epsilon_{i j}\frac{\partial^{j}}{\nabla^{2}}\delta^{2}(x-y), \\[5pt]
\{A_{i} \;,\; \pi^{j}\}_{D_{2}} = \frac{1}{2}\left(\delta_{i}^{j} - \frac{\partial^{j}\partial_{i}}{\nabla^{2}}\right)\delta^{2}(x-y), & \{v_{0} \;,\; \pi^{i}\}_{D_{2}} = -\left(\theta m\frac{1}{2 \nabla^{2}} + \frac{1}{2}\right)\partial^{i}\delta^{2}(x-y), \\[5pt]
\{v_{0} \;,\; \tilde{\pi}^{0}\}_{D_{2}} = \frac{1}{2}\delta^{2}(x-y), & \{v_{0} \;,\; \tilde{\pi}^{i}\}_{D_{2}} =  -\frac{m}{2}\epsilon^{i j}\frac{\partial_{j}}{\nabla^{2}}\delta^{2}(x-y),  \\[5pt]
\{v_{i} \;,\; \tilde{\pi}^{j}\}_{D_{2}} = \frac{1}{2}\left(\delta_{i}^{j} - \frac{\partial_{i}\partial^{j}}{\nabla^{2}}\right)\delta^{2}(x- y), & \{v_{i} \;,\; \pi^{0}\}_{D_{2}} = -\frac{1}{2}\partial_{i}\delta^{2}(x-y), \\[5pt]
\{\pi^{0} \;,\; \tilde{\pi}^{i}\}_{D_{2}} = -\frac{1}{4 m} \epsilon^{i j} {\partial_{j}} \delta^{2}\left(x-y \right), & \{\pi^{i} \;,\; \tilde{\pi}^{0}\}_{D_{2}} = \frac{1}{4m}\epsilon^{i j}\partial_{j}\delta^{2}(x-y), \\[5pt]
\{\pi^{i} \;,\; \tilde{\pi}^{j}\}_{D_{2}} = \frac{1}{4}\left(\delta^{i j} - \frac{\partial^{i}\partial^{j}}{\nabla^{2}}\right)\delta^{2}(x-y). & 
\end{array}
\end{matrix}
\end{eqnarray}
These brackets were  not reported in \cite{16, 17}. They can be used for quantization of the theory by using the methods reported in \cite{24}, where a procedure of gauge fixing is developed in the path integral approach. In this manner, our results extend those reported in the literature. 
\section{Conclussions}
A detailed $HJ$ and $GLT$ analysis for higher-order Maxwell-Chern-Simons theory was developed. Regarding the $HJ$ study, with the introduction of auxiliary fields the theory was written as a first-order time derivative Lagrangian, and by means of the null vectors all Hamiltonians were identified. Then with the introduction of the generalized $HJ$ brackets, all unphysical fields were removed. We then constructed a fundamental differential given in terms of the generalized brackets and involutive Hamiltonians. This allowed us to identify the characteristic equations of the theory, where the equations of motion and the gauge transformations were reported. In this manner, we showed that the $HJ$ is an excellent framework for analyzing higher-order systems. \\
On the other hand, from the $GLT$ we report the complete structure of the constraints. We observed that the constraints were obtained in a consistent way, and there was no need to fix their structure by hand, as developed previously in the literature. Additionally, by fixing the gauge the complete structure of the Dirac brackets was presented. Therefore,  our analysis  extend  those  results presented in \cite{16, 17}, where different approaches were used. Finally, the study developed in this paper can be extended to theories with a more extensive structure, such as gravity and string theory. However, all those results are in progress and will be the subject of forthcoming works \cite{19}.  \\     
\section{Appendix: Gauge transformations}
\subsection{$HJ$ formalism}
We start by calculating the characteristic equations from the fundamental differential, which will reveal the symmetries of the theory. Using (\ref{dfg}), we find them to be
\begin{eqnarray}
\nonumber d\xi_{0} &=& v_{0}dt - d\sigma^{2}, \\ \nonumber 
d\xi_{i} &=& v_{i}dt + \partial_{i}d\sigma^{3}, \\ \nonumber 
d\pi^{0} &=& \left[  \frac{1}{2}{\partial_{i}}v^{i} - \frac{1}{2}\nabla^{2}\xi_{0} + \frac{3\theta}{4}\epsilon^{i j}{\partial_{i}}\xi_{j} - \nabla^{2}\tilde{\pi}^{0} + \frac{1}{4m}\epsilon^{i j}\nabla^{2}{\partial_{i}}\xi_{j} + \frac{1}{2}{\partial_{i}}\pi^{i} \right]dt, \\ \nonumber 
d\pi^{i} &=& \left[ -\partial_{j}F^{i j}- \frac{\theta}{2}\epsilon^{i j}v_{j} - \nabla^{2}\tilde{\pi}^{i} \right] dt - \frac{1}{2m}\epsilon^{i j}\partial_{j}d\sigma^{1} + \left[\frac{\theta}{2}\epsilon^{i j}\partial_{j} + \frac{1}{2m}\epsilon^{i j}\nabla^{2}\partial_{j}\right]d\sigma^{3}, \\ \nonumber 
dv_{0}  &=&  \nabla^{2}\xi_{0}dt + d\sigma^{1},  \\ \nonumber 
dv_{i} &=& \left[\frac{1}{2} \nabla^{2}\xi_{i}+ \frac{1}{2}\partial_{i}v_{0} -m\epsilon_{i j}v^{j} +m \epsilon_{i j}\partial ^{j}\xi_{0} + \frac{\theta m}{2}\xi_{i} + m\epsilon_{i j}\pi^{j}\right] dt - \partial_{i}d\sigma^{2},  \\ \nonumber 
d\tilde{\pi}^{0} &=& -\pi^{0}dt, \\ 
d\tilde{\pi}^{i} &=& \left[ \frac{1}{2}v^{i} - \frac{1}{2}\partial^{i}\xi_{0} + \frac{\theta}{4}\epsilon^{i j}\xi_{j}  - \frac{1}{4m}\epsilon^{i j}\partial_{j}v_{0} + \frac{1}{4m}\epsilon^{i j}\nabla^{2}\xi_{j} - \frac{1}{2}\pi^{i} \right] dt.
\end{eqnarray}
The evolution of the dynamical variables with respect to our parameters $\sigma^{i}$ is understood as canonical transformations, with the corresponding hamiltonians $\Gamma^{i}$ as generators \cite{20, 21}. Due to Frobenius' theorem \cite{21}, the transformation with respect to one of these parameters is independent of the evolution along the others. To relate these canonical transformations to the gauge ones we set $dt=0$  \cite{F20}, obtaining 
\begin{eqnarray}
\delta \xi_{0} &=& -\delta \sigma^{2}, \nonumber \\
\delta \xi_{i} &=& \partial_{i} \delta \sigma^{3}, \nonumber \\
\delta \pi^{0} &=& 0, \nonumber \\
\delta \pi^{i} &=& -\frac{1}{2m} \epsilon^{i j} \partial_{j} \delta \sigma^{1} + \left[\frac{\theta}{2} \epsilon^{i j} \partial_{j}+\frac{1}{2m} \epsilon^{i j} \nabla^{2} \partial_{j}\right] \delta \sigma^{3}, \nonumber \\
\delta v_{0} &=& \delta \sigma^{1}, \nonumber \\
\delta v_{i} &=& -\partial_{i} \delta \sigma^{2}, \nonumber \\
\delta \tilde{\pi}^{0} &=& 0, \nonumber \\
\delta \tilde{\pi}^{i} &=& 0.
\label{eq42}
\end{eqnarray}
In $HJ$, to find the gauge transformations it is necessary to see the specific conditions in which (\ref{eq42}) acts into the Lagrangian. Thus, the Lagrangian (\ref{Lag1}) becomes invariant under these transformations if $\delta L =0$. This will result in relations between the parameters $\sigma^{2}, \sigma^3$. The variation of the Lagrangian is
\begin{eqnarray*}
\delta L = \int dt\;d^{2}x\; \left[\frac{\partial \mathcal{L}}{\partial A_{\mu}}\delta A_{\mu} + \frac{\partial \mathcal{L}}{\partial (\partial_{\nu}A_{\mu})}\delta (\partial_{\nu}A_{\mu}) + \frac{\partial \mathcal{L}}{\partial (\partial_{\nu}\partial^{\mu}A_{\mu})}\delta (\partial_{\nu}\partial^{\mu}A_{\mu})\right],
\end{eqnarray*}
here we use $A_{\mu}$ instead of $\xi_{\mu}$ to more easily compare both formalisms. This, up to a total time derivative, is found to be
\begin{eqnarray}
\delta L = \int dt\;d^{2}x\; \left[\theta\epsilon^{\sigma \nu \lambda}\partial_{\nu} A_{\lambda} + \partial_{\rho}F^{\rho \sigma} - \frac{1}{2m}\epsilon^{\sigma \rho \mu}\left(\partial_{0}\partial^{0}\partial_{\rho}A_{\mu}\right) + \frac{1}{m}\epsilon^{\sigma \nu \lambda}\nabla^{2}\partial_{\nu} A_{\lambda}\right] \delta A_{\sigma} = 0.
\label{varia}
\end{eqnarray}
We can combine the first and second equations in (\ref{eq42}) to write the variation of $A_{\sigma}$ as
\begin{eqnarray}
\delta A_{\sigma} = - \delta_{\sigma}^{0}\delta \sigma^{2} + \delta_{\sigma}^{i} \partial_{i} \delta \sigma^{3},
\label{gauge1}
\end{eqnarray}
thus, by using (\ref{gauge1}) into (\ref{varia}) the variation of the action takes the form 
\begin{eqnarray}
\delta L = -\int dt\;d^{2}x\;\left(\theta\epsilon^{i j}\partial_{i} A_{j} + \partial_{i}F^{i 0} - \frac{1}{2m}\epsilon^{i j}\partial_{i}\ddot{A}_{j} + \frac{1}{m}\epsilon^{i j}\nabla^{2}\partial_{i} A_{j}\right)  \left(\delta \sigma^{2} + \partial_{0} \delta \sigma^{3}\right) = 0.
\end{eqnarray}
The theory will  be invariant under (\ref{eq42})  if the parameters $\sigma^{i}$ obey
\begin{eqnarray}
\delta \sigma^{2} = -\partial_{0} \delta \sigma^{3},
\end{eqnarray}
hence, from (\ref{gauge1}) the gauge transformations are given by
\begin{eqnarray}
\delta A_{\mu} = \partial_{\mu} \delta \sigma^{3}.
\label{gauge}
\end{eqnarray}
Additionally, since $v_{\mu} = \dot{A}_{\mu}$, it can be seen that $\delta \sigma^{1} = \partial_{0}\partial_{0} \delta \sigma^{3}$.
\subsection{GLT formalism}
In this section we use Castellani's procedure \cite{17, 22, 23} to obtain the gauge transformations. We start this calculation with the hamiltonian (\ref{Ham1}), the constraints given  in (\ref{first}), and the Dirac brackets (\ref{DiracGLT}). First, we define the gauge generator as
\begin{eqnarray}
G=\int \epsilon_{a} \gamma^{a} d^{2}x, 
\end{eqnarray}
where $\epsilon_{a}$ are the gauge parameters and $a=1,2,3$. This generates infinitesimal gauge transformations on pase space variables, say $F$, through 
\begin{eqnarray}
\delta F=\int \delta\epsilon_{a}(y)\left\{F(x), \gamma^{a}(y)\right\}_{D} d^{2} y.
\label{GLTgenerator}
\end{eqnarray}
In particular, the generator obeys the following equation, called the master equation,
\begin{eqnarray}
\frac{\partial}{\partial t} G+\left\{G, \mathcal{H}_{T}\right\}_{D}=0.
\end{eqnarray}
Where $\mathcal{H}_{T} = \mathcal{H} + u_{a}\gamma^{a}$ is the total hamiltonian. From the algebra of the constraints and the canonical hamiltonian $\mathcal{H}$ we can obtain the structure functions $V_{b}^{a}$, $C_{c}^{a b}$, given by
\begin{eqnarray}
\left\{\mathcal{H}, \gamma^{a}(\mathrm{x})\right\}_{D} &=& \int d^{2}y\; V_{b}^{a}(x, y) \gamma^{b}(y),  \\
\left\{\gamma^{a}(x), \gamma^{b}(y)\right\}_{D} &=& \int d^{2}z\; C_{c}^{a b}(x, y, z) \gamma^{c}(z).
\end{eqnarray}
Using these, the master equation becomes
\begin{eqnarray}
\frac{d \epsilon_{a}(x)}{d t}-\int d^{2}y\; \epsilon_{b}(y) V_{a}^{b}(x, y) - \int d^{2}y\; d^{2}z\; \epsilon_{b}(y) \gamma_{c}(z) C_{a}^{c b}(x, y, z) = 0.
\end{eqnarray}
Since the only non-zero structure functions are
\begin{eqnarray}
V_{2}^{1}&=& -\delta^{2}(x-y) \quad, \quad V_{3}^{2} = -\delta^{2}(x-y), \nonumber
\end{eqnarray}
with all the $C_{c}^{a b}=0$. We obtain the following relations between the generators.
\begin{eqnarray}
\epsilon_{1} &=& \ddot{\epsilon}_{3}, \nonumber \\
\epsilon_{2} &=& -\dot{\epsilon}_{3}.
\label{relationsGLT}
\end{eqnarray}
Therefore, the generator has only one  parameter and can be written as
\begin{eqnarray}
G=\int d^{2}x\; \left(\delta \ddot{\epsilon}_{3} \gamma^{1} - \delta \dot{\epsilon}_{3} \gamma^{2} + \delta \epsilon_{3} \gamma^{3}\right),
\end{eqnarray}
using (\ref{GLTgenerator}) the gauge transformations of the variables are
\begin{eqnarray}
\delta A_{0} &=& \int \delta \epsilon_{2}(y)\left[\delta^{2}(x-y)\right] d^{2}y, \nonumber \\
\delta A_{i} &=& \int \delta \epsilon_{3}(y)\left[\frac{\partial}{\partial y^{i}} \delta^{2}(x-y)\right] d^{2}y, \nonumber \\
\delta \pi^{0} &=& \int 0 d^{2}y, \nonumber \\
\delta \pi^{i} &=& \int \delta \epsilon_{1}(y)\left[\frac{1}{2m}\epsilon^{i j}\frac{\partial}{\partial x^{j}}\delta^{2}(x-y)\right] + \delta \epsilon_{3}(y)\left[-\frac{\theta}{2}\epsilon^{i j}\frac{\partial}{\partial x^{j}}\delta^{2}(x-y) - \frac{1}{2m}\epsilon^{i j}\nabla_{y}^{2}\frac{\partial}{\partial x^{j}} \delta^{2}(x-y)\right] d^{2}y, \nonumber \\
\delta v_{0} &=& \int \delta \epsilon_{1}(y)\left[-\delta^{2}(x-y)\right] d^{2}y, \nonumber \\
\delta v_{i} &=& \int \delta \epsilon_{2}(y)\left[\frac{\partial}{\partial x^{i}} \delta^{2}(x-y)\right] d^{2}y, \nonumber \\
\delta \tilde{\pi}^{0} &=& \int 0 d^{2}y, \nonumber \\
\delta \tilde{\pi}^{i} &=& \int 0 d^{2}y, 
\end{eqnarray}
and by using (\ref{relationsGLT}) the following gauge transformations are found
\begin{eqnarray}
\delta A_{\mu} &=& -\partial_{\mu} \delta \epsilon_{3}, \nonumber \\
\delta \pi^{\mu} &=& \epsilon^{0 \mu j}\left(- \frac{\theta}{2} + \frac{1}{2m} - \frac{1}{2m}\nabla^{2}\right)\partial_{j} \delta \epsilon_{3}, \nonumber \\
\delta v_{\mu} &=& -\partial_{\mu} \delta \dot{\epsilon}_{3}, \nonumber \\
\delta \tilde{\pi}^{\mu} &=& 0.
\end{eqnarray}
By identifying $\sigma^{3} = - \epsilon_{3}$ both formalisms agree (see equations (\ref{gauge}) and (\ref{eq42})).

\end{document}